\documentclass[12pt]{article}

\usepackage{scicite}

\usepackage{times}

\usepackage{nicefrac,textgreek,wasysym}
\usepackage[utf8]{inputenc}
\usepackage{graphicx}
\usepackage{xr}
\newenvironment{sciabstract}{%
\begin{quote} \bf}
{\end{quote}}

\title{The silver route to cuprate analogs\\
\vspace{1cm}
\normalsize Commercial AgF$_ 2$ is shown to mimic the electronic structure of the parent high-T$_c$ superconducting cuprates.}

\author{Jakub Gawraczyński,$^{1,2}$  Dominik Kurzydłowski,$^{1,3}$   Wojciech Gadomski,$^{2}$\\  Zoran Mazej,$^{4}$ Giampiero Ruani,$^5$ Ilaria Bergenti,$^5$ Tomasz Jaroń,$^{1}$\\ Andrew Ozarowski,$^{6}$  Stephen Hill,$^{6,7}$ Piotr J. Leszczyński,$^{1}$  Kamil Tokár,$^{8}$\\  Mariana Derzsi,$^{1}$  Paolo Barone,$^{9}$  Krzysztof Wohlfeld,$^{10}$\\
 José Lorenzana,$^{11\ast}$  Wojciech Grochala$^{1\ast}$\\
\normalsize{ $^{1}$ Center of New Technologies, University of Warsaw,}\\
\normalsize{ Żwirki i Wigury 93, 02089 Warsaw Poland}\\
\normalsize{ $^{2}$ Faculty of Chemistry, University of Warsaw, Pasteur 1, 02093 Warsaw Poland}\\
\normalsize{ $^{3}$ Faculty of Mathematics and Natural Sciences, Cardinal Stefan Wyszyński }\\ 
\normalsize{ University in Warsaw, Wóycickiego
1/3, 01938 Warsaw Poland}\\
\normalsize{ $^{4}$ Department of Inorganic Chemistry and Technology,}\\
\normalsize{  Jožef Stefan Institute, Jamova 39, SI-1000 Ljubljana,
Slovenia}\\
\normalsize{ $^{5}$Institute of Nanostructured Materials, ISMN-CNR, via Gobetti 101, 40129 Bologna, Italy}\\
\normalsize{ $^{6}$ National High Magnetic Field Laboratory, Florida State University,}\\ 
\normalsize{1800 E. Paul Dirac Drive, Tallahassee,
Florida 32310, USA}\\
\normalsize{$^{7}$  Department of Physics, Florida State University,}\\ \normalsize{Tallahassee, Florida 32306, USA}\\
\normalsize{ $^{8}$ Institute of Physics, Slovak Academy of Sciences, Dúbravská cesta 9, Bratislava, Slovakia}\\
\normalsize{ $^{9}$ Superconducting and other Innovative Materials and Devices Institute,}\\
\normalsize{ SPIN-CNR, via Vetoio, 67100 L’Aquila, Italy}\\
\normalsize{$^{10}$   Faculty of Physics, University of Warsaw, Pasteur 5, 02093 Warsaw Poland}\\
\normalsize{$^{11}$  Institute for Complex Systems, ISC-CNR, Dipartimento di Fisica,}\\
\normalsize{ Universit\`a di Roma ``La Sapienza'', Piazzale Aldo Moro 5, 00185 Roma, Italy}\\
\normalsize{$^\ast$To whom correspondence should be addressed;}\\
\normalsize{ E-mail:  jose.lorenzana@cnr.it,w.grochala@cent.uw.edu.pl.}
}

\date{}

\begin{document}

\baselineskip24pt

\maketitle

\begin{sciabstract}
The parent compound of high-$T_c$ superconducting cuprates is a unique Mott
state consisting of layers of spin-\nicefrac{1}{2} ions forming a square lattice and with a record high in-plane antiferromagnetic coupling. Compounds with
similar characteristics have long been searched for. Nickelates and iridates
had been proposed but have not reached a satisfactory similarity. Here we use a combination of experimental and theoretical tools to show that commercial AgF$_2$ is an excellent cuprate analog with remarkably similar electronic parameters to La$_2$CuO$_4$ but larger buckling of planes. Two-magnon Raman scattering reveals a superexchange constant reaching 70\% of that of a typical cuprate. We argue that structures that reduce or eliminate the buckling of the AgF$_2$ planes could have an antiferromagnetic coupling that matches or surpasses the cuprates.
\end{sciabstract}

Cuprates are said to be unique materials in that they combine layers of
 spin-\nicefrac{1}{2} magnetic moments coupled by a record high antiferromagnetic interaction, strong covalence between transition-metal (TM) and ligands and no orbital degeneracy.  Since the discovery of high-$T_c$ cuprate superconductors, there have been several attempts to replicate these characteristics with different TM ions\cite{Norman2016}. One proposal has been to use nickel(I) in place of copper(II)\cite{Anisimov1999,Chaloupka2008}.
  While LaNiO$_2$ is isoelectronic and isostructural %
with the infinite layer parent cuprates it lacks the strong covalent character between the TM and ligand\cite{Lee2004}, and antiferromagnetic order has not been found\cite{Hayward1999,Ikeda2013}. Sr$_2$IrO$_4$ has many similarities with cuprates including a robust antiferromagnetic order of spin-\nicefrac{1}{2} pseudospins. However, the correlated insulator character is much weaker, and spin-orbit coupling effects dominate\cite{Kim2008,Fujiyama2012}.

\begin{figure}[tb]
  \centering
  \includegraphics[width=16 cm]{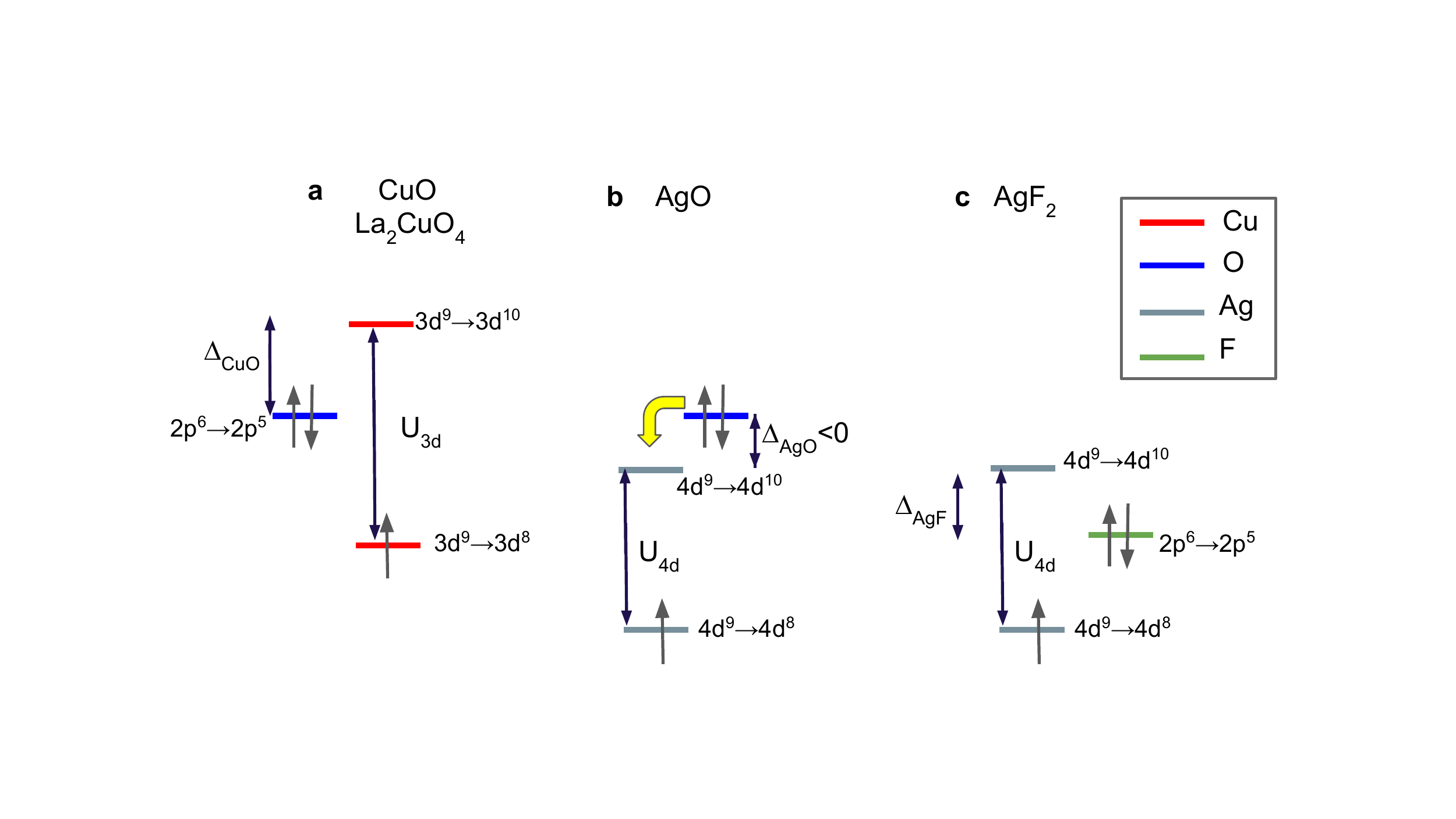} 
  \caption{
{\bf Schematic energy levels of cuprates and argentates in an ionic picture.} 
All levels are assumed to be referenced to a common zero energy vacuum so that the $3d^9\rightarrow 3d^{10}$  ($4d^9\rightarrow 4d^{10}$) level is located at minus the second ionization energy of Cu (Ag). 
{\bf a} Levels for CuO and La$_2$CuO$_4$.  Cu is in the $d^9$ configuration. Electron addition  ($3d^9 \rightarrow 3d^{10}$) and removal
 ($3d^9\rightarrow 3d^{8}$) energies from Cu correspond to the centroid of the upper and lower Hubbard bands while removal from filled oxygen corresponds to the valence band. The charge-transfer energy $\Delta_{CuO}$ and the Hubbard $U_{3d}$ parameter are indicated.  {\bf b} In the case of AgO, silver is formally $d^9$. However, since the $4d^9\rightarrow 4d^{10}$ levels are deeper than the $3d$ counterpart, the charge-transfer energy is practically zero or even negative, and the pictured filling is unstable  towards more complex mixed valence behavior (yellow arrow)\cite{Tjeng1990}. {\bf c} Fluorine is more electronegative than O which translates into deeper $2p^6$ removal states  
and restores a positive charge-transfer energy in AgF$_2$. }
  \label{fig:levels}
\end{figure}

Another obvious possibility is to move down the periodic table from copper to
 silver (Fig.~\ref{fig:levels}). The electronic structure of silver oxides was explored in the early times of high-T$_c$\cite{Tjeng1990}. However, the second ionization potential of silver is nearly 1.2 eV larger than the one of Cu. Therefore, for a compound like AgO  (which is formally $d^9$) it is convenient to transfer electrons from the filled oxygen shells to the TM (Fig.~\ref{fig:levels}{\bf b}). Thus, AgO shows no magnetic ordering as opposed to the formally isoelectronic CuO (Fig.~\ref{fig:levels}{\bf a}). The larger ionization energy of silver can be compensated by a right step in the periodic table replacing oxygen by the more electronegative fluorine (Fig.~\ref{fig:levels}{\bf c}).

Fluoroargentates have been scrutinized some years ago\cite{Grochala2001,Jaron2008}. 
Cs$_2$AgF$_4$ is structurally similar and isoelectronic with La$_2$CuO$_4$ (the so called ``214'' family). However, alternating orbital ordering stabilized by
small structural distortions leads to ferromagnetic intra-sheet interactions 
instead of the robust antiferromagnetic order of cuprates\cite{McLain2006,Grochala2006a}.
Large superexchange constants have been predicted in many fluorargentates\cite{Kurzydowski2017}
and static susceptibility measurements in a quasi-one-dimensional system
suggest a superexchange constant of the order of that found in the cuprates\cite{Kurzydowski2013}.  On the other hand, direct access to excitation energies from spectroscopy has not been available to date.

\begin{figure}[tb]
  \centering
  \includegraphics[width=4 cm]{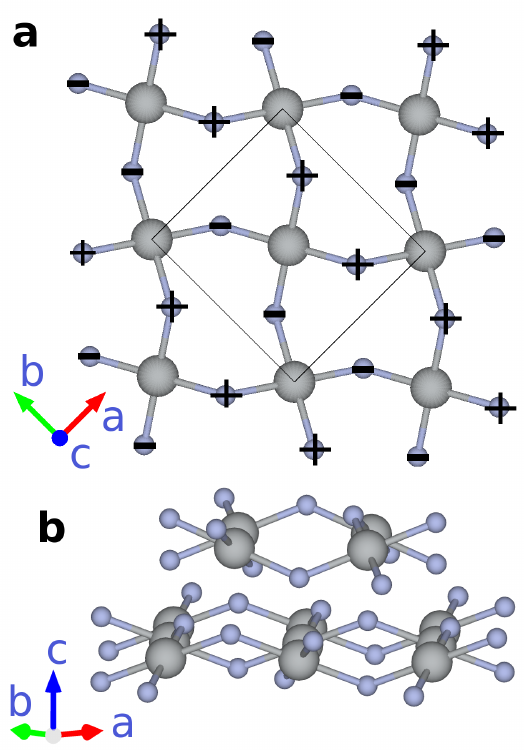} 
  \includegraphics[width=12 cm]{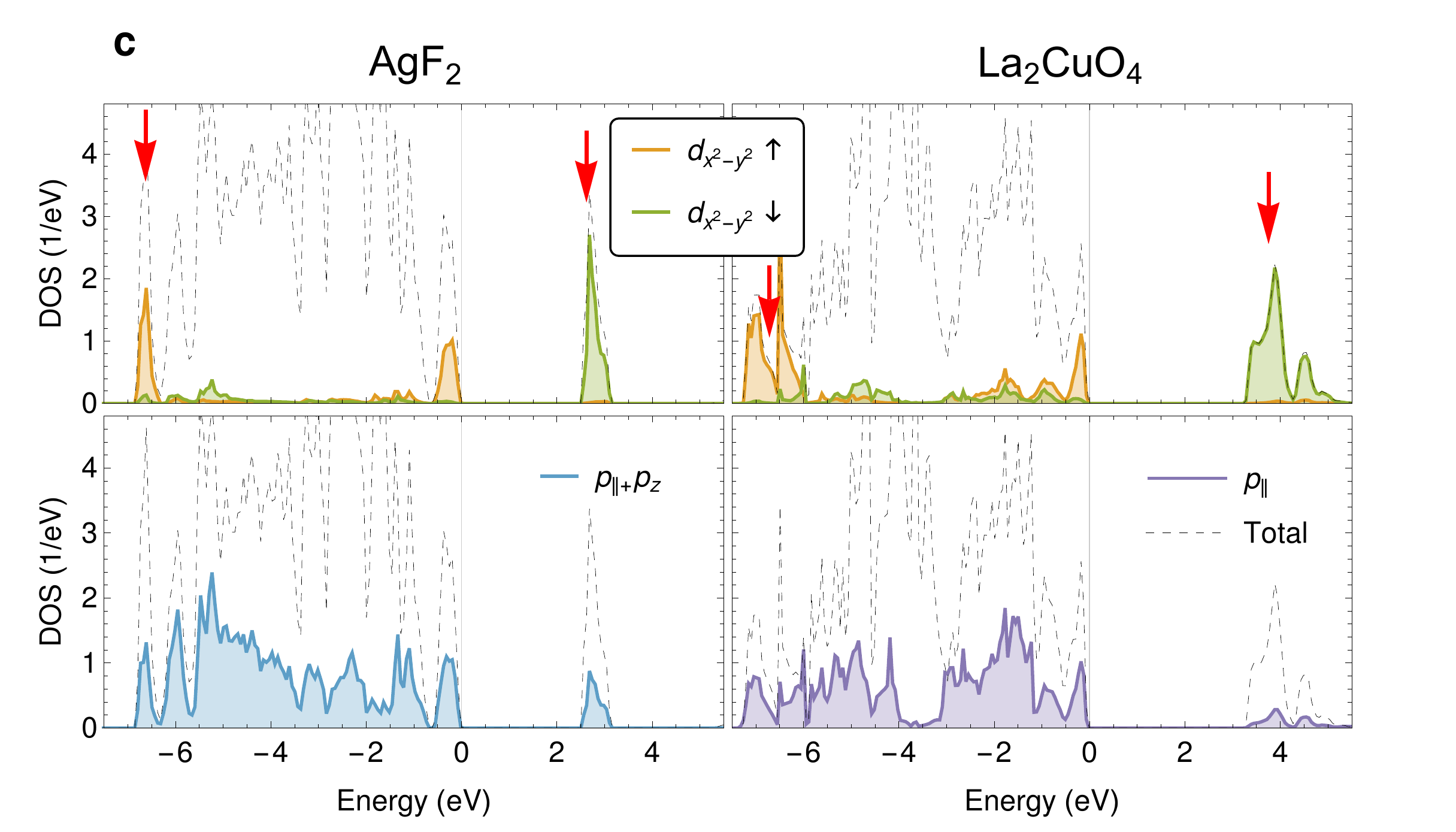} 
  \caption{{\bf  AgF$_2$ vs La$_2$CuO$_4$.}
{\bf a} Top view of a  plane in  AgF$_2$. The black rectangle indicates the unit cell. %
Big/small atoms are Ag/F. We indicate the sign of F displacements in the $c$ direction.  Notice that nearest neighbour exchange paths are equivalent even though  the overall symmetry is orthorhombic as in the LTO phase of La$_2$CuO$_4$. {\bf b} Side view showing buckling and stacking of planes. 
 {\bf c} Comparison of the orbital resolved (upper panels  $d_{x^2-y^2}$, lower panels $p_\parallel$ and $p_z$) 
and total (dashed lines) density of states of the two compounds computed within hybrid-DFT in the antiferromagnetic state. 
The red arrows indicate the Hubbard bands.
}
  \label{fig:agf2}
\end{figure}

A bonus property of F$^{-}$ in place of O$^{2-}$ is that, unlike CuO$_2$ planes,   neutral  AgF$_2$ planes are possible, so the simplest compound is not of the 214 kind but simply 012. Unexpectedly, as we show here, the binary and commercially available AgF$_2$ compound turns out to be an excellent cuprate analog.  Figure~\ref{fig:agf2}{\bf a},{\bf b}  and Supplementary Fig.~\ref{fig:structure} show the structure\cite{Fischer1971} and stacking of planes in  AgF$_2$.  The topology is the same as in  La$_2$CuO$_4$  which, in the low-temperature orthorhombic phase (LTO), has also the same pattern of displacements of the ligands out of the plane (+ or -) but with a much smaller magnitude. Also, as in parent cuprates,  the ground state of AgF$_2$  is antiferromagnetic\cite{Fischer1971} with a staggered moment of 0.7 $\mu_B$ (0.6 $\mu_B$), a weak ferromagnetic component  of $1\times 10^{-2} \mu_B$ ($2 \times 10^{-3}\mu_B$) per 
TM ion and a N\'eel  temperature of $163$K ($325$K), where parentheses  enclose reference values\cite{Kastner1998} in La$_2$CuO$_4$.  

Density functional theory (DFT) computations show that the electronic structure of AgF$_2$ is very similar to that in cuprates.  Fig.~\ref{fig:agf2}{\bf c} shows hybrid-DFT calculations comparing the character of the density of states (see also Ref.~\citenum{Mazej2016}).  Both compounds have well separated Hubbard bands of dominant $d_{x^2-y^2}$ character (indicated by red arrows)  while the first ionization states are of dominant $p$ character, so both compounds are predicted to be charge-transfer insulators according to the Zaanen-Sawatzky-Allen\cite{Zaanen1985} classification scheme. 
The small charge-transfer gap is considered a key characteristic of cuprates. Our unbiased computations in Fig.~\ref{fig:agf2}{\bf c} predict that AgF$_2$ should have an even smaller charge-transfer energy than the cuprates which can provide a test platform for theories based on the smallness of this parameter\cite{Varma1997}.

As a minimal model of the electronic structure we consider 
 one $d_{x^2-y^2}$ orbital centered on the TM (as in cuprates\cite{Emery1987}) and two $p$ orbitals on the F-sites (one more than the cuprates).
 $p_\parallel$  indicates the $p$-orbital parallel to the TM-TM bond, while $p_z$ is perpendicular to the bond but also mixes with the $d_{x^2-y^2}$-orbitals  because of the substantial buckling (see Methods and Supplementary Fig.~\ref{fig:bond}). 
From unpolarized DFT computations we find that the hopping integral of an hypothetically straight Ag-F-Ag bond  with the same interatomic distances 
is $t_{pd}=1.38$ eV,  practically the same value as the cuprates\cite{McMahan1990}  (see Supplementary Figs.~\ref{fig:unpolband} and \ref{fig:tpd}, and Methods). However, due to the increased corrugation of the planes, and the destructive interference of the  $p_\parallel$ and $p_z$ orbitals, the effective hybridization is smaller,  which explains the narrower UHB of AgF$_2$ in Fig.~\ref{fig:agf2}{\bf c} (see Supplementary Table~\ref{tab:tpd}).

The similar splitting between Hubbard bands shown in Fig.~\ref{fig:agf2}{\bf c} for AgF$_2$ and  La$_2$CuO$_4$  (9.4 eV, 10.7 eV  respectively) suggests that
the Hubbard $U_{d}$ parametrizing the Coulomb repulsion on the  $d_{x^2-y^2}$ orbitals is similar in the two compounds. However, this estimate does not take properly into account the  polarizability of the environment.  An empirical estimate of $U_d$ can be obtained from Auger experiments in compounds with similar ions but a filled $d$ shell. Thus, for Cu$_2$O (Cu $d^{10}$) Sawatzky and collaborators\cite{Ghijsen1988} obtained $U_{3d}=9.2$~eV, which is close to the accepted value for Cu $d^{9}$  in CuO$_2$ planes. For  Ag$_2$O (Ag $d^{10}$) they obtained  $U_{4d}=5.8$ eV which is smaller than for $3d$, as expected for the  more diffuse $4d$ orbitals but still quite large. This value, however,  can not be directly transposed to AgF$_2$ because the difference in screening  has to be taken into account.  Screening of the free-ion Hubbard $U_0$ to the 
value $U$ in the solid is determined by the relaxation energy $R$ of the environment according to $U=U_0-R$. For TM compounds, $R$ is expected to scale with the polarization of the ligands and to be inversely proportional to the ligand-TM distance to the fourth power\cite{DeBoer1984}. For copper oxides typical values of the oxygen polarizability are in the range $ \alpha_O =1.9 \sim 3.2$ \AA$^{3}$. Instead,  in AgF$_2$, the ligand-TM distance is larger and the polarizability of the ligand is much smaller $\alpha_F=0.64$ \AA$^{3}$ leading to less efficient screening. This allows for values of $U_d$ in AgF$_2$ similar to the cuprates (see Methods). Interestingly, from a different perspective, the computations of Fig.~\ref{fig:agf2}{\bf c} suggest the same conclusion.

The similarity of the DFT electronic structures confirms the conclusions of the 
simplistic picture of Fig.~\ref{fig:levels}{\bf a} and {\bf c}. From the point of view of the strength of the correlation, 
the narrower upper Hubbard band of AgF$_2$ suggests a more correlated system, but  the similar or smaller $U_{d}$ and smaller charge-transfer gap point in the opposite direction.
Our present results imply that AgF$_2$ is very covalent, 
probably more so than the cuprates,
 which is consistent with core and valence level spectroscopy results\cite{Grochala2003} but in contrast with the prevailing view that fluorides are ionic compounds.

The more interesting parameter for a cuprate analogue is the magnetic exchange interaction which determines the scale of magnetic fluctuations in magnetically mediated mechanisms of superconductivity in doped compounds. Not surprisingly, 
AgF$_2$ is antiferromagnetic\cite{Fischer1971} as predicted by Anderson superexchange.  
A perturbation theory analysis shows that the buckling tends to reduce the superexchange interaction, both because of the reduction of the direct hybridization of $p_\parallel$ with the $d_{x^2-y^2}$ orbitals, and because of the destructive interference of the  $p_z$ orbital (see Methods). 
Unfortunately, the present lack of precise parameters does not allow to obtain an accurate value of the magnetic interactions with this method. 
An alternative estimate can be obtained from hybrid-DFT computations
which yield $J=52$ meV for the nearest-neighbor interaction (see Methods and Ref.~\citenum{Kurzydowski2017}).  This value is 50\% of the typical values in cuprates.

As in La$_2$CuO$_4$, the TM in one plane coincides with the center of the four TM plaquette in the next plane which tends to frustrate interplanar magnetic interactions in an antiferromagnetically ordered state. In both compounds, the frustration is partially relieved by the orthorhombicity but the effect is much larger in AgF$_2$. For the ratio of interlayer to intralayer couplings our DFT computations predict values of the order of  $10^{-2}$ (See Supplementary Table~\ref{tab:j}), a larger ratio than the experimental\cite{Kastner1998} one in cuprates ($10^{-5}$)  but still well in the regime of a quasi two-dimensional quantum antiferromagnet. 

We used the full set of exchange constants from the hybrid-DFT computations to estimate the N\'eel temperature with classical Monte Carlo computations (see Methods). 
Correcting approximately for quantum effects, 
we found a value higher than the experimental one (by 27\%) but of the same
 order. Given the approximations involved, this fair agreement validates the hybrid-DFT computations of exchange constants. 
As shown below, Raman experiments reveal an even larger value of the effective nearest-neighbor exchange. We believe 
uncertainties in the interplanar coupling and extra terms in the Hamiltonian
like four-site cyclic exchange can explain the mild difference in strength between the DFT estimates and Raman (see Methods).

In order to further test the analogy we have experimentally studied 
high quality powder samples of freshly prepared AgF$_2$  (see Methods).  Crystallinity was verified with powder X-ray diffraction which resulted in a spectrum with similar characteristics but less impurity signatures than a commercial sample (see Methods and Supplementary Fig.~\ref{fig:xray}).

 Specific heat measurements reveal that only about 5 \% of the maximum possible entropy change  $R\ln(2)$ occurs around the Ne\'el temperature (see Methods and Supplementary Fig.~\ref{fig:specificheat}). This is consistent with a quasi two-dimensional antiferromagnet where short range in-plane correlations set in at much higher temperatures than the three-dimensional N\'eel temperature.   
An electron pramagnetic resonance signal was searched for over a wide range of temperatures. However,  despite several efforts, it was not found. This is also analogous to the situation in the cuprates and points to strong antiferromagnetic interactions (see Methods). 
Vibrational spectroscopy (Raman and infrared) shows phonon modes in good agreement with the prediction of hybrid-DFT, validating the latter computations  (Supplementary Fig.~\ref{fig:vibration}). In addition, the detailed phonon assignment (see Supplementary Methods and Supplementary Table 1) allowed us to check the integrity of the sample under the laser spot and to exclude impurity phases or photochemistry byproducts .

\begin{figure}[tb]
  \centering
  \includegraphics[width=7.3 cm]{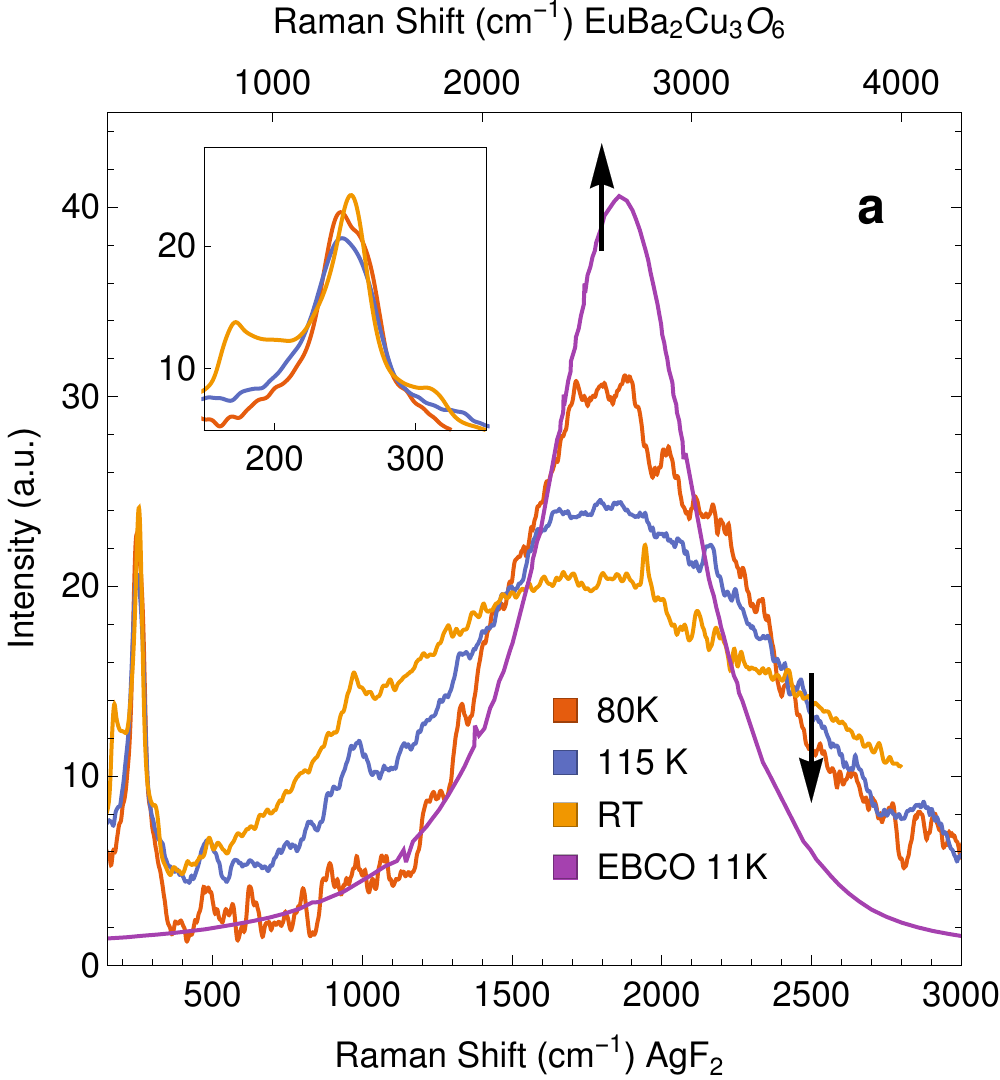}
\includegraphics[width=7. cm]{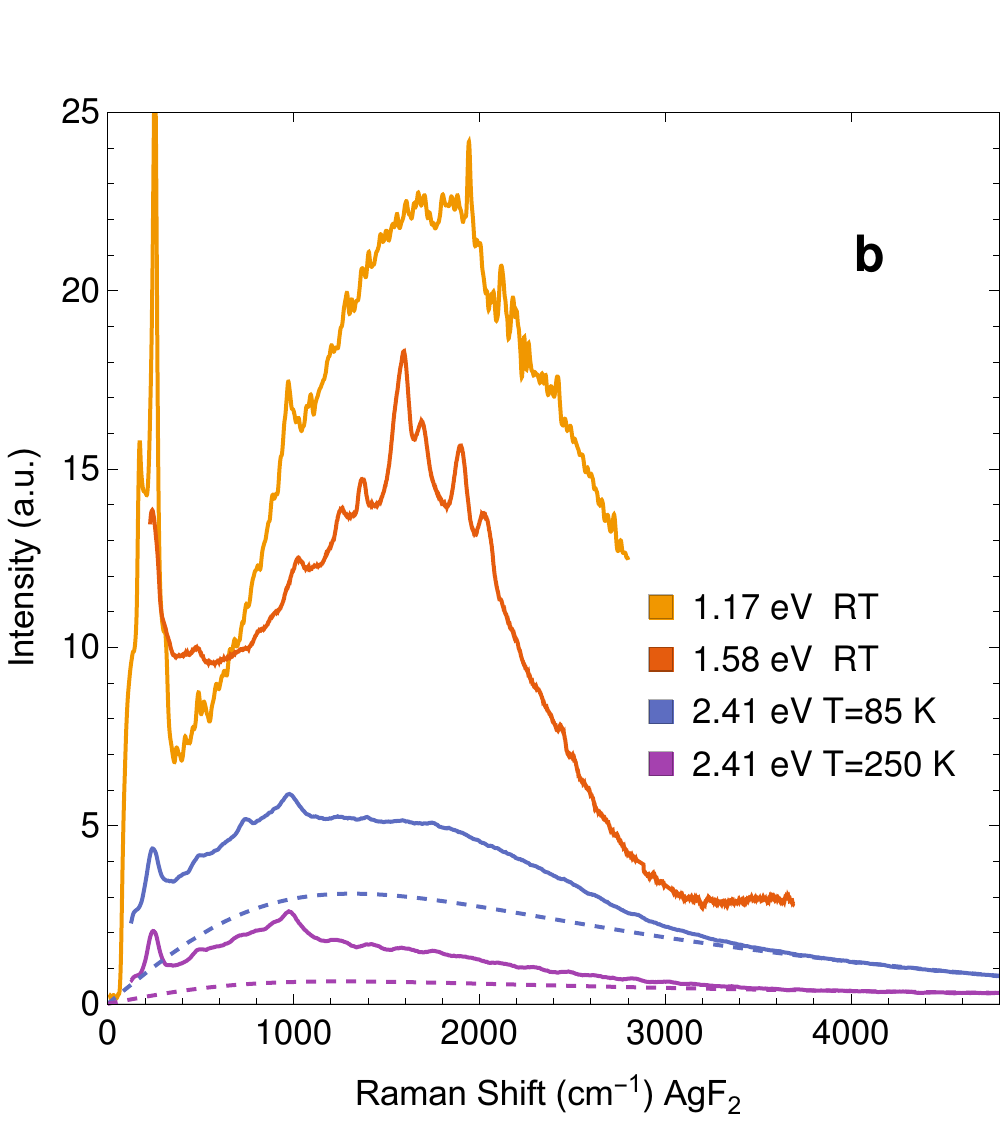}
  \caption{{\bf Two-magnon Raman spectra of  AgF$_2$ } {\bf a}
Temperature dependence for excitation energy  1.17 eV. We also show the 
low temperature two-magnon line shape for the parent cuprate compound EuBa$_2$Cu$_3$O$_6$ (EBCO) after Ref.~\citenum{Knoll1990}. The upper  scale corresponds to EBCO while the lower scale corresponds to AgF$_2$ (present work). Curves are labeled by the temperature in Kelvin.  Comparing the energy scales in {\bf a} we obtain that $J_{AF}=0.7 J_{EBCO}$ which yields  $J_{AF}=70$~meV for AgF$_2$. A linear background was subtracted to the spectra at 80K and 115K but not the the RT spectrum which was measured using a different machine (see Methods). Spectra at different temperatures/apparatus was normalized with the phonon lines as shown in the inset. {\bf b} We show data for AgF$_2$ at various excitation energies.  For the  2.41 eV laser line 
two temperatures are reported. Dashed lines are estimated non-magnetic backgrounds which are well defined at high energy but can not be uniquely defined at low energy. 
 }
  \label{fig:raman}
\end{figure}

For high Raman shifts, we detected a broad band centered at 1750 cm$^{-1}$ at room temperature which hardens and becomes narrower on cooling  Fig.~\ref{fig:raman}{\bf a}. 
The temperature dependence is very similar to the one for two-magnon Raman scattering in cuprates\cite{Knoll1990}. This allows to identify unambiguously this line as due to two-magnon Raman scattering in AgF$_2$. 
Comparing with cuprates we deduce that the antiferromagnetic exchange in this compound is $J_{AF}=70$ meV, confirming
the expected large antiferromagnetic interaction. The robustness of two-magnon Raman scattering well above the N\'eel temperature confirms also strong short range antiferromagentic correlations, also like in cuprates.   

 Fig.~\ref{fig:raman}{\bf b} shows the two-magnon Raman line recorded with three different apparatuses (see Methods). The line shape is clearly visible with an excitation energy of 1.17 eV  and 1.58 eV. The feature at 970 cm$^{-1}$ is assigned to a two-phonon process. For 2.41 eV the two-magnon feature is weak at high temperature (compared to the two-phonon feature) but becomes stronger upon cooling, similar to the behavior for 1.17 eV (panel {\bf a}) and for cuprates. Unfortunately a detailed line shape analysis is not possible in this case due to a strong energy dependent background (dashed lines). 
The dependence on the excitation energy suggest that the resonance profile of the two-magnon line is shifted to lower laser excitation energies with respect to what is found in cuprates\cite{Blumberg1996}, which is consistent with the smaller charge-transfer gap 
found in DFT.  

The above results clearly indicate that AgF$_2$ is an excellent cuprate analog with a very similar electronic structure.  Historically, the superconducting temperature in cuprates has been optimized starting from simple ternary compounds and  increasing the number of chemical elements. In argentates, 
starting from a binary compound that is already stable leaves substantial freedom for such optimizations.  Clearly, a high-priority is to produce 
doped AgF$_2$ planes.  This should allow studying how antiferromagnetism disappears, what is the role of the electron-phonon interaction\cite{Zaanen1994},  and which other states of the correlated electrons arise (pseudogap, charge-density wave, superconductivity, etc.).

The cuprate family of parent high-T$_c$ superconductors is made of materials very similar to each other. Indeed, changes in key parameters like $J$ are smaller than 15\%\cite{Tokura1990}. This severely hampers the identification of clear trends in physical properties. A close cuprate analog but with important differences (less marked two-dimensional character, smaller $J$, narrower bands, smaller charge-transfer gap) opens the way to the clarification of the mysterious cuprate phase diagram by revealing clear trends.
Another priority is to produce compounds with flat AgF$_2$ planes. We have estimated using DFT that such compounds should have an antiferromagnetic $J$ which is more than twice the one of commercial AgF$_2$ (see Methods and Supplementary Table~\ref{tab:jpoly}) and, assuming a magnetically driven mechanism\cite{Scalapino2012a},  could potentially lead to superconducting critical temperatures higher than those exhibited by cuprates.

\bibliography{scibib,library}

\begin{thebibliography}{10}

\bibitem{Norman2016}
M.~R. Norman, {Materials Design for New Superconductors}, {\it Reports Prog.
  Phys.\/} {\bf 79}, 074502 (2016).

\bibitem{Anisimov1999}
V.~I. Anisimov, D.~Bukhvalov, T.~M. Rice, {Electronic structure of possible
  nickelate analogs to the cuprates}, {\it Phys. Rev. B\/} {\bf 59}, 7901
  (1999).

\bibitem{Chaloupka2008}
J.~Chaloupka, G.~Khaliullin, {Orbital order and possible superconductivity in
  LaNiO$_3$/LaMO$_3$ superlattices}, {\it Phys. Rev. Lett.\/} {\bf 100}, 3
  (2008).

\bibitem{Lee2004}
K.~W. Lee, W.~E. Pickett, {Infinite-layer LaNiO$_2$: Ni$^{1+}$ is not
  Cu$^{2+}$}, {\it Phys. Rev. B - Condens. Matter Mater. Phys.\/} {\bf 70}, 1
  (2004).

\bibitem{Hayward1999}
M.~A. Hayward, M.~A. Green, M.~J. Rosseinsky, J.~Sloan, {Sodium hydride as a
  powerful reducing agent for topotactic oxide deintercalation: Synthesis and
  characterization of the nickel(I) oxide LaNiO$_2$}, {\it J. Am. Chem. Soc.\/}
  {\bf 121}, 8843 (1999).

\bibitem{Ikeda2013}
A.~Ikeda, T.~Manabe, M.~Naito, {Improved conductivity of infinite-layer
  LaNiO$_2$ thin films by metal organic decomposition}, {\it Phys. C Supercond.
  its Appl.\/} {\bf 495}, 134 (2013).

\bibitem{Kim2008}
B.~J. Kim, {\it et~al.\/}, {Novel $J_{eff}=1/2$ Mott state induced by
  relativistic spin-orbit coupling in Sr$_2$IrO$_4$}, {\it Phys. Rev. Lett.\/}
  {\bf 101}, 1 (2008).

\bibitem{Fujiyama2012}
S.~Fujiyama, H.~Ohsumi, T.~Komesu, J.~Matsuno, B.~J. Kim, M.~Takata, T.~Arima,
  H.~Takagi, {Two-dimensional Heisenberg behavior of $J_{eff}=1/2$ isospins in
  the paramagnetic state of the spin-orbital Mott insulator Sr$_2$IrO$_4$},
  {\it Phys. Rev. Lett.\/} {\bf 108}, 1 (2012).

\bibitem{Tjeng1990}
L.~H. Tjeng, M.~B. J.~B. Meinders, J.~van Elp, J.~Ghijsen, G.~A. Sawatzky,
  R.~L. Johnson, {Electronic structure of Ag$_2$O}, {\it Phys. Rev. B\/} {\bf
  41}, 3190 (1990).

\bibitem{Grochala2001}
W.~Grochala, R.~Hoffmann, {Real and Hypothetical Intermediate-Valence
  AgII/AgIII and AgII/AgI Fluoride Systems as Potential Superconductors}, {\it
  Angew. Chemie Int. Ed.\/} {\bf 40}, 2742 (2001).

\bibitem{Jaron2008}
T.~Jaro{\'{n}}, W.~Grochala, {Prediction of giant antiferromagnetic coupling in
  exotic fluorides of Ag}, {\it Phys. Status Solidi - Rapid Res. Lett.\/} {\bf
  2}, 71 (2008).

\bibitem{McLain2006}
S.~E. McLain, M.~R. Dolgos, D.~A. Tennant, J.~F.~C. Turner, T.~Barnes,
  T.~Proffen, B.~C. Sales, R.~I. Bewley, {Magnetic behaviour of layered Ag(II)
  fluorides.}, {\it Nat. Mater.\/} {\bf 5}, 561 (2006).

\bibitem{Grochala2006a}
W.~Grochala, {Magnetism: Small changes, big consequences}, {\it Nat. Mater.\/}
  {\bf 5}, 513 (2006).

\bibitem{Kurzydowski2017}
D.~Kurzyd{\l}owski, W.~Grochala, {Prediction of Extremely Strong
  Antiferromagnetic Superexchange in Silver(II) Fluorides: Challenging the
  Oxocuprates(II)}, {\it Angew. Chemie - Int. Ed.\/} {\bf 56}, 10114 (2017).

\bibitem{Kurzydowski2013}
D.~Kurzyd{\l}owski, Z.~Mazej, Z.~Jagli{\v{c}}i{\'{c}}, Y.~Filinchuk,
  W.~Grochala, {Structural transition and unusually strong antiferromagnetic
  superexchange coupling in perovskite KAgF$_3$}, {\it Chem. Commun.\/} {\bf
  49}, 6262 (2013).

\bibitem{Fischer1971}
P.~Fischer, G.~Roult, D.~Schwarzenbach, {Crystal and magnetic structure of
  silver difluoride-II. Weak 4d-ferromagnetism of AgF$_2$}, {\it J. Phys. Chem.
  Solids\/} {\bf 32}, 1641 (1971).

\bibitem{Kastner1998}
M.~A. Kastner, R.~J. Birgeneau, G.~Shirane, Y.~Endoh, {Magnetic, transport, and
  optical properties of monolayer copper oxides}, {\it Rev. Mod. Phys.\/} {\bf
  70}, 897 (1998).

\bibitem{Mazej2016}
Z.~Mazej, D.~Kurzyd{\l}owski, W.~Grochala, {\it {Unique Silver(II) Fluorides:
  The Emerging Electronic and Magnetic Materials}\/} (Elsevier, 2016).

\bibitem{Zaanen1985}
J.~Zaanen, G.~A. Sawatzky, J.~W. Allen, {Band gaps and electronic structure of
  transition-metal compounds}, {\it Phys. Rev. Lett.\/} {\bf 55}, 418 (1985).

\bibitem{Varma1997}
C.~M. Varma, {Non-Fermi-liquid states and pairing instability of a general
  model of copper oxide metals }, {\it Phys. Rev. B\/} {\bf 55}, 14554 (1997).

\bibitem{Emery1987}
V.~J. Emery, {Theory of high-Tc superconductivity in oxides}, {\it Phys. Rev.
  Lett.\/} {\bf 58}, 2794 (1987).

\bibitem{McMahan1990}
A.~K. McMahan, J.~F. Annett, R.~M. Martin, {Cuprate parameters from numerical
  Wannier functions}, {\it Phys. Rev. B\/} {\bf 42}, 6268 (1990).

\bibitem{Ghijsen1988}
J.~Ghijsen, L.~Tjeng, J.~van Elp, H.~Eskes, J.~Westerink, G.~Sawatzky,
  M.~Czyzyk, {Electronic structure of {Cu$_2$O} and CuO}, {\it Phys. Rev. B\/}
  {\bf 38}, 11322 (1988).

\bibitem{DeBoer1984}
D.~K.~G. {De Boer}, C.~Haas, G.~A. Sawatzky, {Exciton satellites in
  photoelectron spectra}, {\it Phys. Rev. B\/} {\bf 29}, 4401 (1984).

\bibitem{Grochala2003}
W.~Grochala, R.~G. Egdell, P.~P. Edwards, Z.~Mazej, B.~{\v{Z}}emva, {On the
  covalency of silver-fluorine bonds in compounds of silver(I), silver(II) and
  silver(III)}, {\it ChemPhysChem\/} {\bf 4}, 997 (2003).

\bibitem{Knoll1990}
P.~Knoll, C.~Thomsen, M.~Cardona, P.~Murugaraj, {Temperature-dependent lifetime
  of spin excitations in RBa$_2$Cu$_3$O$_6$ ( R =Eu, Y)}, {\it Phys. Rev. B\/}
  {\bf 42}, 4842 (1990).

\bibitem{Blumberg1996}
G.~Blumberg, P.~Abbamonte, M.~V. Klein, W.~C. Lee, D.~M. Ginsberg, L.~L.
  Miller, A.~Zibold, {Resonant two-magnon Raman scattering in cuprate
  antiferromagnetic insulators}, {\it Phys. Rev. B\/} {\bf 53}, R11930 (1996).

\bibitem{Zaanen1994}
J.~Zaanen, P.~B. Littlewood, {Freezing electronic correlations by polaronic
  instabilities in doped La$_2$NiO$_4$}, {\it Phys. Rev. B\/} {\bf 50}, 7222
  (1994).

\bibitem{Tokura1990}
Y.~Tokura, S.~Koshihara, T.~Arima, H.~Takagi, S.~Ishibashi, T.~Ido, S.~Uchida,
  {Cu-O network dependence of optical charge-transfer gaps and spin-pair
  excitations in single-CuO$_2$-layer compounds}, {\it Phys. Rev. B\/} {\bf
  41}, 11657 (1990).

\bibitem{Scalapino2012a}
D.~J. Scalapino, {A common thread: The pairing interaction for unconventional
  superconductors}, {\it Rev. Mod. Phys.\/} {\bf 84}, 1383 (2012).

\end{thebibliography}

\bibliographystyle{Science_title_10names}

\section*{Acknowledgments}

WG would like to thank to A. Michota-Kaminska for making Renishaw inVia Raman system available. We acknowledge support from the Polish National Science Centre (NCN) within the HARMONIA project “HP” (project 2012/06/M/ST5/00344). Z.M. gratefully acknowledges the Slovenian Research Agency (ARRS) for the financial support of the present study within research program P1-0045 Inorganic Chemistry and Technology. The high-field EPR spectra were recorded at the NHMFL, which is funded by the NSF through cooperative agreement no. DMR-1157490 and the State of Florida. S.H. thanks the NSF for financial support (grant DMR-1610226). Research at the University of Warsaw was carried out with the use of CePT infrastructure financed by the European Union – the European Regional Development Fund within the Operational Program "Innovative economy" for 2007-2013 (POIG.02.02.00-14-024/08-00).
The quantum-mechanical calculations performed in Warsaw would not be possible without the ICM supercomputer Okeanos (ADVANCE PLUS, GA67-13). 
J.L. acknowledges support from a CINECA  high-performance computer  project (IsC50\_HTCM) and from Italian MAECI under collaborative Projects SUPERTOP-PGR04879 and  AR17MO7.

\section*{Supplementary materials}
Materials and Methods\\
Figs. S1 to S7\\
Tables S1 to S4\\
References \textit{(31-75)}

\end{document}